\documentclass[prl, reprint, superscriptaddress, showpacs, bibliography]{revtex4-1}
\usepackage[dvips]{graphicx}
\usepackage{color}
\usepackage{amsmath}
\usepackage{amssymb}
\usepackage[breaklinks]{hyperref}
\usepackage[all]{hypcap}

\begin{document}

\title{Detection of a Disorder-Induced Bose-Einstein Condensate\\ in a Quantum Spin Material at High Magnetic Fields}

\author{A. Orlova}
\affiliation{Laboratoire National des Champs Magn\'etiques Intenses, LNCMI-CNRS (UPR3228), \\
EMFL, UGA, UPS, and INSA, Bo\^{i}te Postale 166, 38042, Grenoble Cedex 9, France}

\author{H. Mayaffre}
\affiliation{Laboratoire National des Champs Magn\'etiques Intenses, LNCMI-CNRS (UPR3228), \\
EMFL, UGA, UPS, and INSA, Bo\^{i}te Postale 166, 38042, Grenoble Cedex 9, France}

\author{S. Kr\"{a}mer}
\affiliation{Laboratoire National des Champs Magn\'etiques Intenses, LNCMI-CNRS (UPR3228), \\
EMFL, UGA, UPS, and INSA, Bo\^{i}te Postale 166, 38042, Grenoble Cedex 9, France}

\author{M. Dupont}
\affiliation{Laboratoire de Physique Th\'{e}orique, IRSAMC, Universit\'{e} de Toulouse, CNRS, UPS, 31062 Toulouse, France}

\author{S. Capponi}
\affiliation{Laboratoire de Physique Th\'{e}orique, IRSAMC, Universit\'{e} de Toulouse, CNRS, UPS, 31062 Toulouse, France}

\author{N. Laflorencie}
\email{laflo@irsamc.ups-tlse.fr}
\affiliation{Laboratoire de Physique Th\'{e}orique, IRSAMC, Universit\'{e} de Toulouse, CNRS, UPS, 31062 Toulouse, France}

\author{A. Paduan-Filho}
\affiliation{Instituto de F\'isica, Universidade de S{\~a}o Paulo, 05315-970 S{\~a}o Paulo, Brazil}

\author{M. Horvati\'{c}}
\email{mladen.horvatic@lncmi.cnrs.fr}
\affiliation{Laboratoire National des Champs Magn\'etiques Intenses, LNCMI-CNRS (UPR3228), \\
EMFL, UGA, UPS, and INSA, Bo\^{i}te Postale 166, 38042, Grenoble Cedex 9, France}

\begin{abstract}
The coupled spin-1 chains material NiCl$_2$-4SC(NH$_2$)$_2$ (DTN) doped with Br impurities is expected to be a perfect candidate for observing many-body localization at high magnetic field: the so-called ``Bose glass'', a zero-temperature bosonic fluid, compressible, gapless, incoherent, and short-range correlated. Using nuclear magnetic resonance (NMR), we critically address the stability of the Bose glass in doped DTN, and find that it hosts a novel disorder-induced \emph{ordered} state of matter, where many-body physics leads to an unexpected resurgence of quantum coherence emerging from localized impurity states. An experimental phase diagram of this new ``order-from-disorder'' phase, established from NMR $T_1^{-1}$ relaxation rate data in the (13 $\pm$ 1)\% Br-doped DTN, is found to be in excellent agreement with the theoretical prediction from large-scale quantum Monte Carlo simulations.
\end{abstract}

\date{\today}

\maketitle

Understanding the subtle effect of disorder in quantum interacting systems is one of the major challenges of modern condensed matter physics. The presence of random impurities or defects in regular crystalline materials breaks the symmetry of translation, and may lead to novel physical phenomena, in particular at a very low temperature where quantum effects are dominant. A very famous example is the Anderson localization \cite{anderson_absence_1958,abrahams_scaling_1979} in the absence of interaction, where quantum interference of electronic waves due to multiple scattering processes induced by the impurities can completely block the transport, thus driving a metal-to-insulator phase transition. Extending this prediction towards realistic condensed matter systems, where inter-particle interactions cannot be ignored, is a highly non-trivial issue which is hard to track experimentally, with only a few examples: metal-insulator transition in two-dimensional silicon MOSFETs \cite{kravchenko_metalinsulator_2004}, localization of ultra-cold atoms in quasi-periodic potentials \cite{derrico_observation_2014,schreiber_observation_2015,choi_exploring_2016}, Cooper pairs localization in disordered superconducting thin films \cite{sacepe_localization_2011}.

Interestingly, there is a set of condensed matter quantum systems for which the interplay between disorder and interactions can be investigated in details: the so-called antiferromagnetic Mott insulators, where low-energy physics is governed by spin degrees of freedom. In such systems, the amount of disorder can be controlled by chemical doping, in contrast with other types of materials where intrinsic disorder is unavoidably present but more difficult to quantify and control.
Among this family of quantum antiferromagnets, there is a large class of systems where the topology of the microscopic interactions between quantum spins is such that the non-magnetic ground state is separated from the first excited magnetized state by a finite energy gap, above which the excitations can proliferate. Upon applying a sufficiently strong external magnetic field $H \geq H_c$, this gap can be closed, and the quantum dynamics of excitations can be faithfully described by effective bosons \cite{mila_ladders_1998,giamarchi_coupled_1999}, thus leading to non-trivial bosonic states of matters, such as Bose-Einstein condensates (BEC) \cite{giamarchi_boseeinstein_2008} or frustration-induced bosonic crystals, as seen, for instance, in the Shastry-Sutherland material Sr$_2$Cu(BO$_3$)$_2$ \cite{kodama_magnetic_2002}.
There are several examples of quantum spin systems hosting a BEC phase \cite{zapf_bose-einstein_2014}: coupled dimers in TlCuCl$_3$ \cite{nikuni2000,ruegg_boseeinstein_2003}, spin-ladder materials such as CuBr$_4$(C$_5$H$_{12}$N)$_2$ (BPCB) \cite{klanjsek_controlling_2008} and CuBr$_4$(C$_7$H$_{10}$N)$_2$ (DIMPY) \cite{jeong_attractive_2013},  bilayer system BaCuSi$_2$O$_6$ (Han purple) \cite{jaime_magnetic-field-induced_2004,sebastian_dimensional_2006,kramer_spatially_2013}.

\begin{figure*}[t!]
\begin{center}
\includegraphics[width=0.8\textwidth,clip]{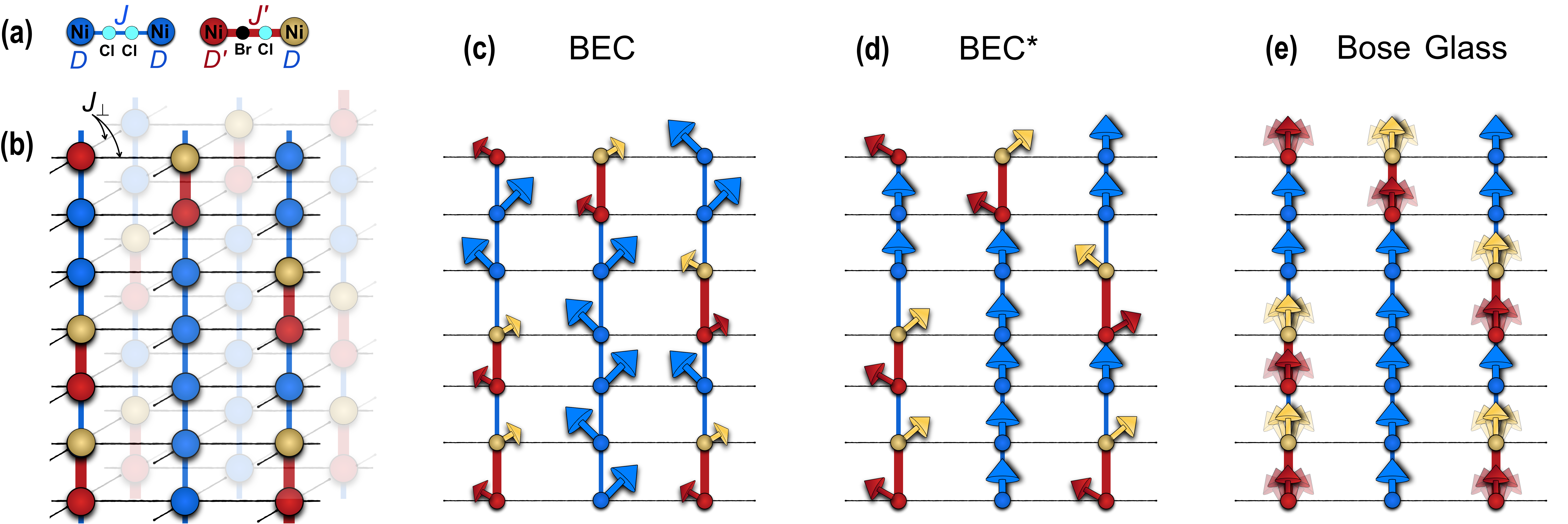}
\caption{(a) Doping Br to replace Cl atoms modifies locally both the affected bond, $J \to J'$, and the single-ion anisotropy of the nearest spin, $D \to D'$, denoted respectively by red bonds and dots in the three-dimensional structure shown in (b). (c) The canted spin polarization in the BEC phase is marginally perturbed by doping. (d) In the BEC* phase, the order is formed among partially polarized doped sites, on a fully polarized background of regular spins. (e) In the Bose-glass regime, the canted polarization at doped sites is uncorrelated and fluctuating. The $c$-axis direction is horizontal in (a) and vertical in (b-e).}
\label{cartoon}
\end{center}
\end{figure*}

In this study, we focus on one of the most convenient and the best known representatives of the field-induced BEC, the NiCl$_2$-4SC(NH$_2$)$_2$ (DTN) quantum antiferromagnet \cite{zapf_bose-einstein_2006,zvyagin2007,wulf2015,blinder_nuclear_2017}. This three-dimensional system, schematically represented in Figs. \ref{cartoon}(a) and \ref{cartoon}(b) and described by the following $S=1$ Hamiltonian
\begin{eqnarray}
    {\cal{H}} &=& \sum_i \Big \{
    \sum_n \Bigl[ J\,{\boldsymbol{S}}_{i,n}\cdot {\boldsymbol{S}}_{i+1,n} + D\left(S_{i,n}^{z}\right)^2-g\mu_B H S_{i,n}^z \Bigr]\nonumber\\
    &+& \sum_{\langle n\,m\rangle}J_\perp\,{\boldsymbol{S}}_{i,n}\cdot {\boldsymbol{S}}_{i,m}
     \Big \},
    \label{eq:DTNX}
\end{eqnarray}
consists of weakly coupled chains of $S = 1$ spins, borne by Ni$^{2+}$ ions, subject to an easy-plane anisotropy $D = 8.9$ K and a nearest-neighbor Heisenberg interaction, dominant along the chain ($c$-axis direction), $J = 2.2$ K, and much weaker from one chain to another, $J_{\perp} = 0.18$ K \cite{zvyagin2007}. The magnetic field $H$ is applied along the $c$-axis direction, which is (by the crystal symmetry) the hard axis of the single-ion anisotropy.

\begin{figure*}[t!]
\begin{center}
\includegraphics[width=0.8\textwidth,clip]{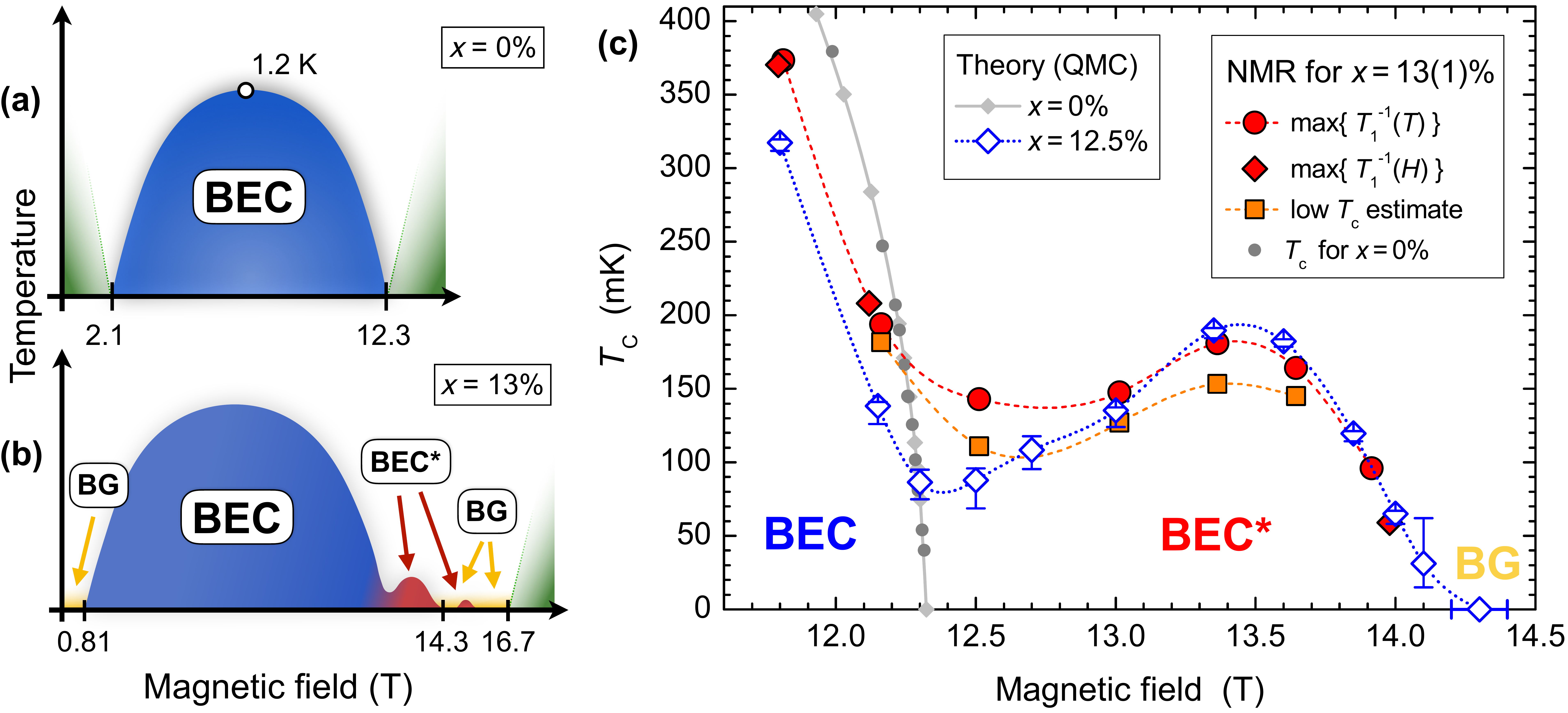}
\caption{Sketch of the global phase diagram of (a) pure DTN and (b) 13\% doped DTN$X$, where colors denote the BEC (blue) and BEC* (red) phases, and the Bose-glass (BG, yellow) and gapped (green) regimes. (c) Focus on the main BEC* phase. The $T_c$ determined from QMC simulations for 12.5\% doping (blue open diamonds) is compared to $T_c$ estimates from the $T_1^{-1}$ NMR data in a (13 $\pm$ 1)\% doped sample that are shown in Fig. \ref{T1}: solid red dots and diamonds denote, respectively, the maximum of $T_1^{-1}(T)$ and $T_1^{-1}(H)$ dependence, reflecting the maximum of the critical spin fluctuations, while orange squares provide the lowest estimate for the $T_c$, taken to be the point where the $T_1^{-1}(T)$ data turn into their BEC regime (see the text). The gray small dots and diamonds are, respectively, the experimental points and the QMC simulation of the BEC phase boundary in the pure DTN, as reported in \cite{blinder_nuclear_2017}. Lines are guide to the eyes.
}
\label{phase_diagram}
\end{center}
\end{figure*}

In the ideal clean case where disorder can be safely ignored, this material displays a BEC for magnetic field $H$ between $H_{c1} = 2.1$ T and $H_{c2} = 12.3$ T [Fig. \ref{phase_diagram}(a)], showing critical properties of the BEC transition, {\it{e.g.}}, a critical temperature $T_c \propto |H - H_{c1,2}|^{2/3}$ \cite{blinder_nuclear_2017} and a striking $\lambda$ anomaly in the specific heat \cite{zapf_bose-einstein_2006}. The BEC phase is characterized by the development of the canted spin polarization, where its component that is transverse to the magnetic field is antiferromagnetically ordered [Fig. \ref{cartoon}(c)].

The situation becomes even more interesting when this material is doped with impurities that locally modify the antiferromagnetic exchanges. Indeed, upon doping with bromine,  Ni(Cl$_{1-x}$Br$_x$)$_2$-4SC(NH$_2$)$_2$ (DTN$X$) provides a fascinating realization of coupled $S=1$ chains with randomness in their couplings, the amount of disorder being chemically controlled by the bromine concentration $x$. As previously determined using NMR \cite{orlova_nuclear_2017}, Br impurities change the affected coupling, $J \to J'\cong 5.3$ K, and the nearest single-ion anisotropy, $D \to D'\cong 3.2$ K [Figs. \ref{cartoon}(a) and \ref{cartoon}(b)]. A direct consequence of this substantial change in $J$ and $D$ is that the gap is locally modified, as first noted by Yu {\it{et al.}} \cite{yu_bose_2012}. Upon applying an external magnetic field, the local magnetization on the perturbed bonds is then reduced as compared to their clean (chlorine only) counterpart [Fig. \ref{cartoon}(c)].  At a sufficiently strong field, above $H_{c2}$, while unperturbed bonds are already totally polarized by the field, the doped bonds can remain partially polarized [Figs. \ref{cartoon}(d) and \ref{cartoon}(e)]. This spin density depolarization (with respect to full polarization) is faithfully described by a two-level system that can be seen as a hard-core boson degree of freedom \cite{orlova_nuclear_2017,dupont_disorder-induced_2017}. Remarkably, these effective bosons cannot easily delocalize over the lattice as they are strongly trapped by the gapped (polarized) ferromagnetic background. We are therefore left with a collection of \textit{a priori} localized bosons whose density is controlled by the magnetic field in the range $H_{c2} < H < H_{c3}$, where $H_{c3} \cong 16.7$ T is the upper critical field required to completely polarize the disordered system.

In their seminal work on DTN$X$,  Yu {\it{et al.}} \cite{yu_bose_2012} concluded that these effective bosons do realize a Bose glass \cite{Giamarchi1988,fisher_boson_1989}, an elusive state of matter which is compressible, gapless, short-range correlated, with \emph{no} phase coherence even at zero temperature: in short, a many-body localized ground-state [Fig. \ref{cartoon}(e)]. However, many-body effects at play in DTN$X$ should be regarded more carefully, as suggested by Refs. \cite{orlova_nuclear_2017,dupont_disorder-induced_2017,dupont_competing_2017}. Indeed, while single-particle states are clearly localized in the close vicinity of perturbed bonds, as demonstrated from microscopic measurements by NMR \cite{orlova_nuclear_2017}, interactions are expected to completely change the picture, inducing \emph{a many-body delocalization}. More precisely, exact diagonalization calculations \cite{orlova_nuclear_2017} combined with large scale quantum Monte Carlo (QMC) simulations \cite{dupont_disorder-induced_2017,dupont_competing_2017} performed on the realistic disordered Hamiltonian given by Eq. (1) have predicted that these localized states should experience an effective unfrustrated \cite{Comment} pairwise coupling, leading to a many-body state with a revival of phase coherence [Fig. \ref{cartoon}(d)], in stark contrast with previously reported Bose-glass physics \cite{yu_bose_2012}. Unlike the original BEC state in DTN, which is only weakly affected by doping, this new ``BEC*'' state is created by impurities and is thus strongly inhomogeneous, yet fully coherent (delocalized). Detailed theoretical investigation of the doping dependence \cite{dupont_competing_2017} has uncovered a rather complex phase diagram of DTN$X$ above $H_{c2}$ (see Fig.~2 in Ref.~\cite{dupont_competing_2017}), where at least three BEC* phases [two of which are visible in Fig.~\ref{phase_diagram}(b)] emerge in the previously expected Bose-glass regime at low doping. These BEC* phases broaden with doping to merge together and finally completely replace the Bose glass at high doping.

On the experimental side, the preceding NMR experiments \cite{orlova_nuclear_2017}, that brought a basis for the microscopic description of the doped sites, have been carried out (for technical convenience) on relatively lightly, 4\% doped sample, where the transition into the BEC* phase is theoretically predicted to be well below 40 mK \cite{dupont_disorder-induced_2017}, and could not be reached by the employed dilution refrigerator. As by stronger doping the predicted phase diagram is pushed up into the experimentally accessible range of temperature, we investigated by $^{14}$N NMR the (13 $\pm$ 1)\% doped DTN$X$ in order to check for the existence of the disorder-induced BEC* phase and establish its experimental phase diagram (Fig. \ref{phase_diagram}). In general, a second order phase transition into an antiferromagnetically ordered phase can be detected by NMR using either static or dynamic observables. In the former case, we directly observe the growth of the order parameter, here the transverse spin polarization, through the splitting or broadening of NMR lines. While this is nicely visible in pure DTN \cite{blinder_nuclear_2017}, in a strongly doped DTN$X$ sample this method could not be employed, because the NMR lines are much broader and the relevant signal much weaker. The point is that the order of the BEC* phase is established only for the minority sites, those affected by doping [Fig. \ref{cartoon}(d)].

\begin{figure*}[t!]
\begin{center}
\includegraphics[width=0.8\textwidth,clip]{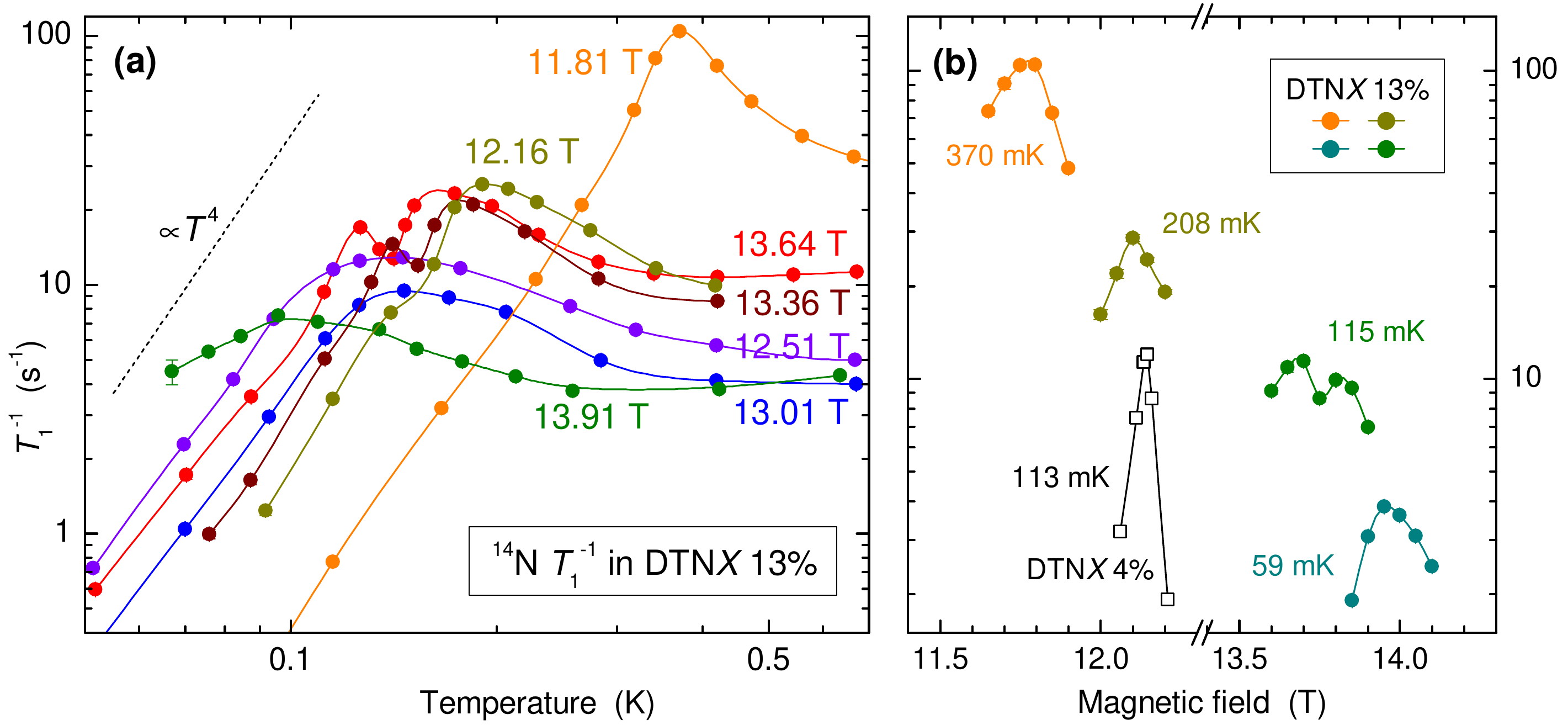}
\caption{ (a) Temperature dependence of $T_1^{-1}$ in 13\% doped DTN$X$ at selected field values. The strong decrease of $T_1^{-1}$ at low temperature is a signature of the ordered BEC phase, while the relatively broad maximum corresponds to the critical spin fluctuations at the phase transition into this phase. The error bars are less than the symbol size, except for one point.
(b) Magnetic field dependence of $T_1^{-1}$ in the vicinity of the phase transition into the ordered phase. Several positions of the 13\% doped DTN$X$ phase diagram (solid dots) are compared to the case of much less, 4\% doped DTN$X$ (open squares), to show how strongly is the $T_1^{-1}$ peak broadened by increasing the doping-induced disorder.}
\label{T1}
\end{center}
\end{figure*}

Fortunately, the ordering temperature $T_c$ can also be determined from the corresponding peak of critical spin fluctuations, measured through the nuclear spin-lattice relaxation rate, $T_1^{-1}$. Indeed, in a pure or weakly disordered system, a sharp peak in the observed $T$ or $H$ dependence of $T_1^{-1}(T,H)$ data precisely defines the $T_c$. However, this peak is broadened by disorder, and in the 13\% doped DTN$X$ data shown in Fig. \ref{T1} only a very broad maximum of $T_1^{-1}$ is separating the high temperature regime above the BEC* phase, where the relaxation is essentially constant, from the rapid power-law decrease, $T_1^{-1} \propto T^4$, observed inside this phase. This latter behavior is the same in the BEC [see the 11.81~T curve in Fig. \ref{T1}(a)] and the BEC* phases, confirming the common nature of these two phases. The high value of the power-law exponent is only slightly smaller than what is usually observed in BEC phases of quantum antiferromagnets (5.0-5.5) \cite{Mayaffre2000,Jeong2017}, reflecting a high-order relaxation process \cite{Beeman1968}. While this low temperature behavior of $T_1^{-1}(T)$ is clearly a fingerprint of the ordered phase, the exact experimental determination of $T_c$ is not evident. The position of the maximum in $T_1^{-1}(T)$ or $T_1^{-1}(H)$, shown in Fig. \ref{phase_diagram}(c), is a first guess, but we can wonder if in a strongly disordered and thus inhomogeneous system the point where the full 3D coherence is established is not somewhat shifted away from the maximum of spin fluctuations, probably towards lower temperature. The lower estimate of $T_c$ is then given by the point where the $T_1^{-1}(T)$ dependence switches into its power-law regime of the ordered phase. We here applied the most simple definition of this point, as given by the crossing of the corresponding ``power-law line'' [in the log-log plot of Fig. \ref{T1}(a)] with the horizontal (constant) line passing through the maximum of $T_1^{-1}(T)$.

In Figs. \ref{T1}(a) and \ref{T1}(b), we can also notice that for some magnetic field values (in both $T$ and $H$ dependences), close to the $T_1^{-1}$ maximum, there is a small dip in the relaxation rate. We observed an anomalous change of the long-time behavior in the corresponding relaxation curves, and suspect that this is due to a crosstalk between the faster relaxing N(1) site used for $T_1^{-1}$ measurement and the other, much slower relaxing N(2) site \cite{blinder_nuclear_2017}. This crosstalk might be in some cases enhanced at the real $T_c$ position, as the position of the dip is very close to our lower $T_c$ estimate.

Fig. \ref{phase_diagram}(c) remarkably shows that thus obtained experimental estimates of the 13\% doped DTN$X$ phase boundary are qualitatively identical and quantitatively quite close to the theoretical QMC predictions, in particular regarding the maximum $T_c$ value of the BEC* phase and its upper field boundary. The observed differences are minor: in the theoretical prediction, the BEC part of the ordered phase is pushed by 0.1~T towards the lower field and the minimum of $T_c$ between the BEC and the BEC* part is (therefore) significantly deeper. The NMR data thus provide the final proof for the existence of the new disorder-induced phase, whose microscopic nature is revealed by the corresponding, precisely defined theoretical description \cite{dupont_disorder-induced_2017,dupont_competing_2017}. In particular, theory tells us that the BEC* phase, in contrast to a Bose glass, is indeed fully 3D coherent. NMR, being a \emph{local} microscopic technique, cannot directly provide this information, but the BEC* phase is accessible to neutron experiments, which can provide further insights on how the coherence is established in this highly inhomogeneous system.

As in our strongly doped DTN$X$ sample the weakly perturbed BEC and strongly perturbed BEC* phases are continuously connected [Figs. \ref{phase_diagram}(b) and \ref{phase_diagram}(c)], the effective level of disorder notably varies along the phase boundary. This is explicitly visible in the magnetic field dependence of the width and the size of the $T_1^{-1}(T)$ peak that reflects the critical spin fluctuations [Fig. \ref{T1}(a)]. The BEC* phase in DTN$X$ thus appears to be a remarkable model to investigate how a phase transition is modified by strong disorder, calling for further experimental and theoretical investigation. For example, here we might mention the so-called ``$\phi$ crisis'' \cite{yu_bose_2012,yao2014} regarding the theoretically predicted $\phi \ge 2$ exponent ensuring smooth critical behavior of the $T_c(H) \propto |H - H_c|^{\phi}$ phase boundary between superfluid and Bose-glass phases in disordered systems \cite{yu_bose_2012,yu2012b,yao2014}. In quantum spin systems, experiments and numerical simulations rather find exponents varying from the expected $\phi = 2/3$ in pure 3D systems to $\phi \simeq 1$ in disordered systems \cite{yu_bose_2012}. However, the predicted upper-field BEC* to Bose glass boundary at $H_{\rm BG} \cong 14.3$ T shown in Fig. \ref{phase_diagram}(c) seems to be more compatible with $\phi > 1$, as expected from theory, although the numerical error bars prevent a definite conclusion.

This new type of inhomogeneous ordered state of matter that we have detected both theoretically and experimentally belongs to the fascinating problem of order-from-disorder phases, first discussed by Villain {\it{et al.}} \cite{Villain1980}. Several fundamental aspects remain to be understood, for instance, how the \emph{inhomogeneity} of the order parameter in the BEC$^*$ phase develops when approaching the transition to the many-body localized Bose glass. Furthermore, as the doping dependence of the $T$--$H$ phase diagram of DTN$X$ is now well understood, we can focus on the narrow windows in which the Bose-glass phase does exist, to study its not yet well understood finite-$T$ properties or excitations. One appealing issue concerns the possible existence of a finite-temperature glass transition above the Bose-glass regime, where, despite the absence of a coherent response, the transverse susceptibility is expected to diverge \cite{fisher_boson_1989} with an unknown critical exponent.

\begin{acknowledgments}
We acknowledge support of the French ANR project BOLODISS (Grant No. ANR-14-CE32-0018) and R\'egion Midi-Pyr\'en\'ees. The numerical work was performed using HPC resources from GENCI (Grants No. x2016050225 and No. x2017050225) and CALMIP. A. P.-F. acknowledges support from the Brazilian agencies CNPq and FAPESP (Grant No. 2015-16191-5).
\end{acknowledgments}


\begin{thebibliography}{99}
\bibitem{Comment}
Unfrustrated here means that for any relative positions of two impurities, their coupling keeps the same antiferromagnetic ``$(-1)^{i+j+k}$'' character of the original pure system.
\end{thebibliography}
\end{document}